\documentclass[prc,aps,showpacs]{revtex4}
\usepackage{graphicx}
\begin{document}

\title{Effect of exotic S=+1 resonances on $K^0_L p$ scattering data}
\author{R.~L.~Workman}
\email{rworkman@gwu.edu}
\author{R.~A.~Arndt}
\author{I.~I.~Strakovsky}
\affiliation{Department of Physics, The George Washington 
             University, Washington, D.C. 20052-0001}
\author{D.~M.~Manley}
\email{manley@kent.edu}
\author{J.~Tulpan}
\affiliation{Department of Physics, Kent State University, 
Kent, OH 44242-0001} 

\begin{abstract}

We consider the effect of an exotic S=+1 $\Theta^+$ resonance on
the scattering of neutral kaons off protons. Explicit results are
presented for the $K^0_L p$ total cross sections. 

\end{abstract}


\maketitle

Results from a wide range of recent experiments are now consistent with the
existence of an exotic S=+1 resonance, the $\Theta^+(1540)$, 
having a narrow width and a mass near 1540~MeV~\cite{Nakano}. 
Width determinations have been hindered by limitations 
on experimental resolutions, resulting in upper bounds of
order 10~MeV. The quantum numbers of this state remain unknown,
though a prediction of $J^P$ = $1/2^+$ was obtained in the
work~\cite{DPP} that provided a motivation for the original search.

An examination of older $K^+d$ and $K^+p$ data has provided
no confirmation for the $\Theta^+$ or an associated 
$\Theta^{++}$ state~\cite{Page}. 
In fact, these older measurements seem to
require a width much smaller than 10~MeV~\cite{Nussinov,DAC,HK,CaT}.
Most investigations of the older data have focused on $K^+d$
experiments from which the $K^+ n$ interaction has been extracted. 
The effect of Fermi motion in the deuteron is particularly
important for a narrow structure, making the observation
of a `bump' in any cross section unlikely. This point has
been extensively demonstrated by Nussinov~\cite{Nussinov}, 
and by Cahn and Trilling~\cite{CaT}.

The problems of Fermi motion can be avoided if instead one
considers the $K_L^0 p$ interaction. However, as the 
$K^0_L$ is a mixture of $K^0$ and $\overline{K^0}$, this 
approach requires an accounting of the $\overline{K}N$ 
interaction.  This we have done, combining results from 
analyses of $\overline{K}N$~\cite{Y1} and $K N$~\cite{Z0} 
data.

\begin{figure}[th]
\centering{
\includegraphics[height=0.7\textwidth, angle=90]{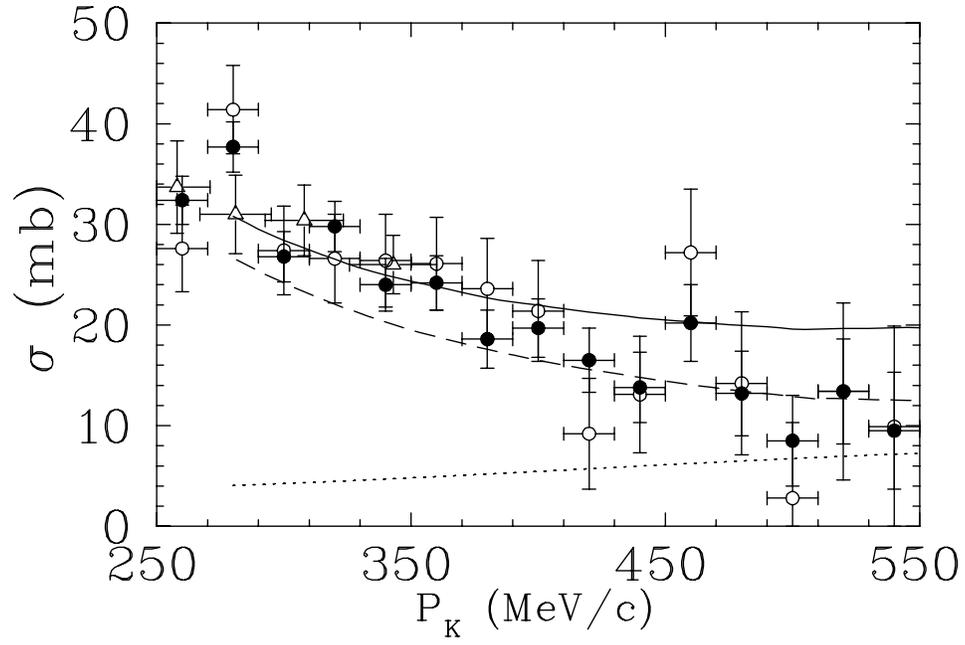}
}\caption{Total cross section for $K^0_L p$ (solid) and 
          contributions (dashed) from S = $-1$~\protect\cite{Y1} 
          and (dotted) S = +1~\protect\cite{Z0}. Data sets 
          from Ref.~\protect\cite{Cleland} (open and solid circles)
          and Ref.~\protect\cite{Sayer} (open triangles).
          \label{fig:g1}}
\end{figure}

The amplitude for $K_L^0 p$ scattering is given by
\begin{equation}
M_{K^0_L p} \; = \; {1\over 4} ( Z_1 + Z_0 + 2 Y_1 )
\end{equation}
where $Z_{0,1}$ are the strangeness 
S = 1, I = 0 and 1 amplitudes, and $Y_1$ is
the S = $-1$, I = 1 amplitude. The $Y_1$ contribution dominates at low
energies, the S = 1 component growing in relative importance with 
increasing energy. Our result for the 
$K^0_L p$ total cross section, calculated from the imaginary part of
the forward scattering amplitude, is given in Fig.~\ref{fig:g1}. 
Note that this is not a fit but rather a prediction
based on analyses of other reactions. 

\begin{figure}[th]
\centering{
\includegraphics[height=0.7\textwidth, angle=90]{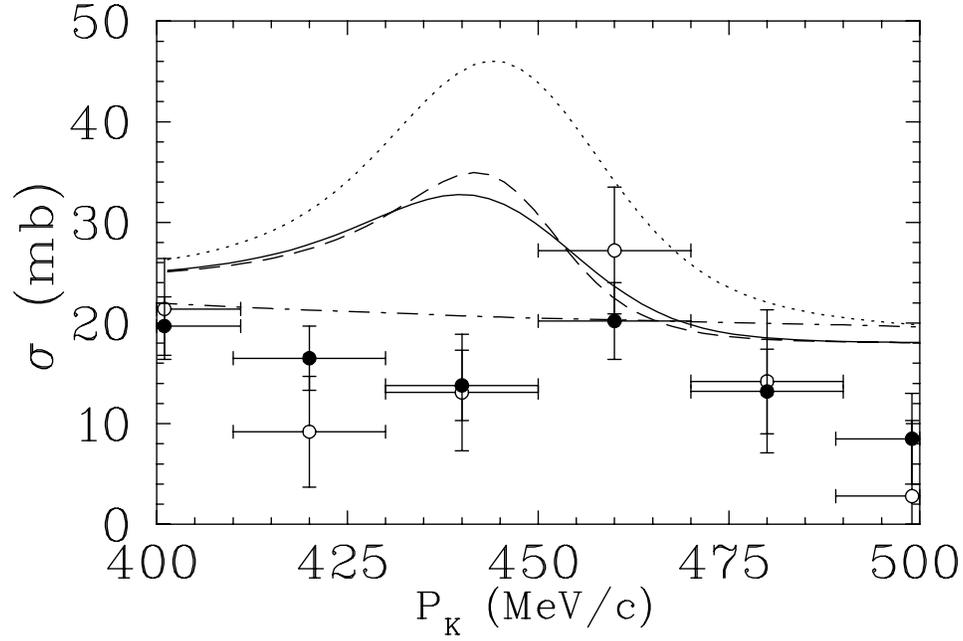}
}\caption{Effect of 1540~MeV $P_{01}$ resonance of width 5~MeV 
          for a 10~MeV/$c$ (dashed) and 20~MeV/$c$ (solid) momentum 
          spread. Effect of a $D_{03}$ resonance of same mass 
          and width (20~MeV/$c$ spread) displayed for comparison 
          (dotted).  Dash-dotted curve gives the unmodified total 
          cross section for $K^0_L p$.  Data as 
          in Fig.~\protect\ref{fig:g1}. \label{fig:g2}}
\end{figure}

Starting from this description of the data, we have added
a narrow Breit-Wigner resonance in order to demonstrate the magnitude
of its effect. In doing so, we have taken into account
the fact that the incident $K^0_L$ beam has a momentum
spread, which also tends to smear out a narrow structure.
In Fig.~\ref{fig:g2}, we have added an S = +1 resonance, 
having a 
5~MeV width, to the $P_{01}$
partial wave. Here we compare the result for beam momentum
distributions (assumed Gaussian) having widths of 10 and 20~MeV/$c$
at 440 MeV/$c$,
the latter being an estimate of the momentum spread
associated with the beam used in Ref.~\cite{Cleland}. 
A resonance in the $D_{03}$ partial wave is also included for 
comparison.

\begin{figure}[th]
\centering{
\includegraphics[height=0.7\textwidth, angle=90]{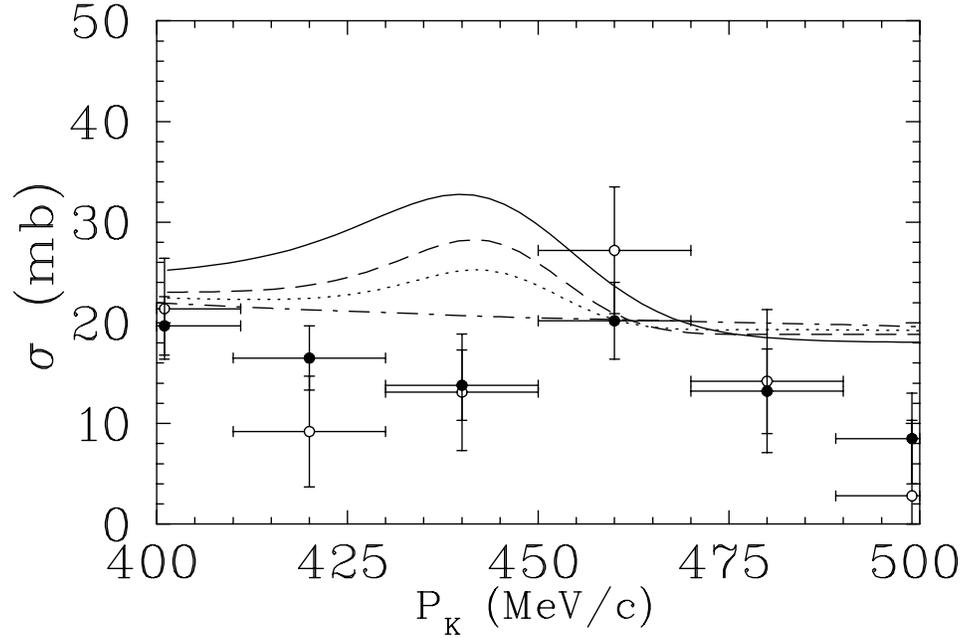}
}\caption{Resonance signal for a 1540~MeV $P_{01}$ resonance 
          of width 1~MeV (dotted), 2~MeV (dashed), and 5~MeV 
          (solid), for a 20~MeV/$c$ momentum spread.  
          Dash-dotted curve gives the unmodified total cross section 
          for $K^0_L p$.  Data as in 
          Fig.~\protect\ref{fig:g1}. \label{fig:g3}}
\end{figure}

The measurement of Ref.~\cite{Cleland} was 
reported with total cross sections calculated over 20~MeV/$c$
bins. We have accordingly averaged over this interval, 
finding
a further minimal reduction in the peak. Results for resonances with
a range of widths are compared in Fig.~\ref{fig:g3}.
From Fig.~\ref{fig:g1}, the most pronounced deviations from 
our predicted smooth
behavior occur near 280 and 460~MeV/$c$ (corresponding to C.M. energies
of 1480 and 1550~MeV). The apparent `bump' at 1480~MeV could be more
than a statistical fluctuation. 
The PDG~\cite{PDG} reports a 1-star $\Sigma (1480)$
based on an analysis of $K^-p\to\overline{K^0}p\pi^-$, 
with a 3.5 standard deviation signal being seen in 
$\overline{K^0}p$. (The data of Ref.\cite{Sayer}
does not show this structure.)
Given the large experimental uncertainties,
a fluctuation near 1550~MeV is only interesting in that it
occurs near the expected $\Theta^+(1540)$ signal. (The overall momentum
scale has a quoted uncertainty of 2\%.) Here too the PDG reports a
weak evidence for a nearby bump, the $\Sigma (1560)$. 

In conclusion, present $K_L^0 p$ scattering data are insufficiently precise to
confirm the $\Theta (1540)$. 
However, if more precise data were to become available, with improved
momentum resolution,
this method would have the advantage of producing a resonance structure,
unlike the $K^+ n$ cross sections extracted from deuteron target 
experiments which are fundamentally limited by Fermi momentum. 
In this case, the main limiting factor for a determination
of $\Theta^+$ properties would be our knowledge of weak
$\Sigma$ resonances\cite{Yakov}.

\acknowledgments

This work was supported in part by the U.~S. Department of 
Energy Grants DE--FG02--99ER41110 and 
DE-FG02-01ER41194.  We acknowledge useful communications 
with Ya.~I.~Azimov and A.~G.~Dolgolenko.
R.~W. and I.~S. gratefully 
acknowledge a contract from Jefferson Lab under which 
this work was done.  Jefferson Lab is operated by the 
Southeastern Universities Research Association under the 
U.~S.~Department of Energy Contract DE--AC05--84ER40150.


\eject


\end{document}